\documentclass[conference,a4paper]{IEEEtran}
\ifCLASSINFOpdf
\else
\fi

\usepackage{graphicx}
\usepackage{epsfig,rotating,latexsym,amsmath,amssymb,setspace,enumerate,subfigure,multirow}
\usepackage{algorithm}
\usepackage{algpseudocode}
\usepackage{cite}
\usepackage{url}
\usepackage[usenames]{color}
\usepackage[normalem]{ulem}
\usepackage{setspace}

\begin{document}
%
\title{Architecting Information Centric ETSI-M2M systems}
%
\author{\IEEEauthorblockN{Luigi Alfredo Grieco\IEEEauthorrefmark{2}\IEEEauthorrefmark{1},
Mahdi Ben Alaya \IEEEauthorrefmark{1},
Thierry Monteil \IEEEauthorrefmark{1} and
Khalil Drira\IEEEauthorrefmark{1}}
\IEEEauthorblockA{\IEEEauthorrefmark{1}LAAS-CNRS and Univ. de Toulouse, Toulouse, France. \IEEEauthorrefmark{2}DEI, Politecnico di Bari, Bari, Italy.\\
Email: a.grieco@poliba.it; \{grieco, ben.alaya, monteil, khalil\}@laas.fr}
}


\maketitle

\begin{abstract}
The European Telecommunications Standards Institute (ETSI) released a set of specifications to define a restful architecture for enabling seamless service provisioning across heterogeneous Machine-to-Machine (M2M) systems.
The current version of this architecture is strongly centralized, thus requiring new enhancements to its scalability, fault tolerance, and flexibility. To bridge this gap, herein it is presented an Overlay Service Capability Layer, based on Information Centric Networking design. Key features, example use cases and preliminary performance assessments are also discussed to highlight the potential of our approach. 
\end{abstract}


%
\IEEEpeerreviewmaketitle

\section{Introduction}
Machine-to-Machine (M2M) communication technologies promise to inter-connect up to 50 billion devices by 2020, thus paving the way to advanced pervasive applications in several domains \cite{Boswarthick:2012:MCS:2378488}. 
Current M2M market, unfortunately, is very fragmented. Many vertical M2M solutions have been designed independently for different applications, which inevitably hinder a large-scale M2M deployment \cite{interdigital:11}. 


To this end, the European Telecommunications Standards Institute (ETSI) come up with a set of specifications which provides a restful architecture able to standardize the way heterogeneous devices can offer services and access to them seamlessly \cite{Boswarthick:2012:MCS:2378488}. 
The building blocks of an ETSI M2M system are: devices, gateways, and networks. A device is a machine equipped with a set of resources/services that can be made accessible to the rest of the system. Many devices may also bind to the same gateway in order to make their resources visible outside their local domain. Finally, the resources available at many gateways and devices are exposed at a wide area scope through an ETSI M2M network. 
The standard defines a functional architecture with a generic set of service capabilities for M2M on top of connectivity layers deployed in M2M networks, gateways, and devices. In this abstract functional architecture each physical equipment is represented as an instance of a Service Capability Layer (SCL). Each device, gateways, and the network will find a corresponding entity in the SCL, so that the abstract model will be a collection of Device SCL (SCL), Gateway SCL (GSCL), and Network SCL (NSCL). Moreover, each SCL instance is responsible for a subset of resources modeled and named according a recursive hierarchical tree. 

\begin{figure}[htbp]
	\centering
		\includegraphics[width=8 cm]{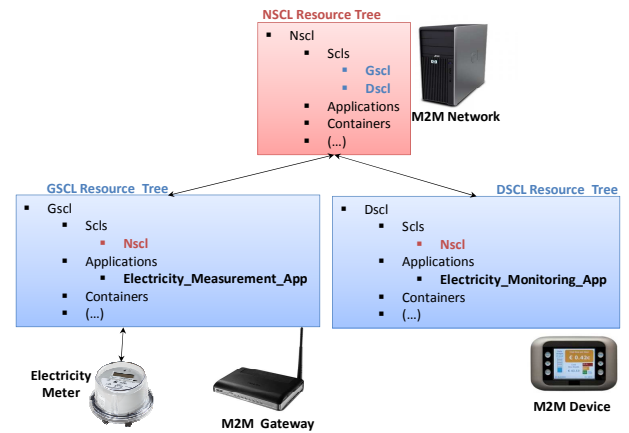}
	\caption{A Simple ETSI M2M scenario.}
	\label{etsi-m2m-example1}	
\end{figure}

At the present stage, the ETSI M2M architecture is very centralized and the  system critically depends on the NSCL (see also Fig. \ref{etsi-m2m-example1}). In fact, the NSCL is in charge of handling: (1) mutual authentications; (2) resource discovery operations;  (3) subscriptions; and (4) data relaying. In other words, in M2M networks with millions of device the NSCL has to be able to take part to all the active data sessions by relaying both signaling information (i.e., subscriptions) and data exchanged between remote endpoints (i.e., devices and gateways). This design choice might impair the scalability of the system and its fault tolerance. 
Furthermore, the overall scale of the system could be limited by a too centralized architecture because each single event that happens in the M2M infrastructure has to be relayed by the NSCL. Moreover, even if the NSCL can be implemented in a distributed way, its scale and costs would be much higher if the capabilities of GSCL and DSCL are not fully exploited.
For these reasons, recent studies \cite{interdigital:11} advocate extensions to the M2M standard that could lead to more scalable and distributed operations based on a meshed overlays of SCL instances. Also the oneM2M standard (which embraces ETSI M2M) is following the same direction \cite{oneM2M}. 
\textit{At the same time, the management of a meshed overlay that connect many SCL instances requires new challenges to afford in order to ensure effective service discovery and access to M2M resources.} This problem becomes even more relevant if we consider that distributed caching mechanisms can be used in ETSI M2M systems.

A satisfactorily answer to such challenges can be found in  the emerging Information Centric Networking (ICN) paradigm, which grounds its foundation on name based content discovery mechanisms and, more in general, on name oriented networking primitives in distributed networks of caches \cite{Xylomenos:2013, Heidemann:2001, Okamoto:2012, Marias:2012, Rayes:2012, Ravindran:2013}. Many ICN architectures are available nowadays, all having the common ambition to move a step beyond the current host centric Internet design towards the Future Internet. Also, ICN architectures are natively capable to embrace, merge, and extend most of the technical solutions conceived in peer-to-peer systems literature so far \cite{Xylomenos:2013}.  
The Information-Centric Networking Research Group (ICNRG) within the Internet Research Task Force (IRTF) is currently defining unifying baseline terminology, evaluation methodology, and design rules for ICN \cite{Kostas:2013a}.

Since the resources of ETSI M2M systems are named according to a recursive hierarchical tree, we propose herein to design an Overlay Service Capability Layer (OSCL) in accordance to the Named Data Networking (NDN) architecture\footnote{www.named-data.org} \cite{Jacobson:09}, which, among the available ICN platforms, is the leading one based on hierarchical names. Moreover, the way NDN has been conceived will allow native support to multicast applications, seamless wireless connectivity, content level security, and in network caching mechanisms, thus improving the scalability, fault tolerance, and flexibility of ETSI M2M systems.
Noticeably, the OSCL we propose could be implemented also in ICN architectures other than NDN, provided that they allow an efficient management of hierarchical content names.

\section{NDN Overview}\label{secndn}
The NDN architecture is based on the Content Centric Networking (CCN) rationale \cite{Jacobson:09} that assumes hierarchical content names, receiver initiated sessions, content level security schema, and in-network caching mechanisms. The NDN architecture is born to natively support mobile applications and multicast data dissemination while, at the same time, to improve content distribution services across heterogeneous platforms.
The entire NDN paradigm is based on two packets: \textit{Interest} and \textit{Data}. When a consumer of data needs a given information it just sends an \textit{Interest} message in which it is specified the hierarchical name of the requested item. This \textit{Interest} message is routed towards one or multiple nodes that are in possess of the requested content, which is then encapsulated in a \textit{Data} message and sent back as an answer to the consumer.  
Along its path to the requester, the \textit{Data} message can be stored in the cache of relaying nodes, that will become able to answer to future requests for the same content, thus lowering the load on the servers that store permanent copies of the asked item. Furthermore, if a NDN router receives multiple requests for the same content that have been issued by different nodes, it will forward only the first \textit{Interest} packet and it will suppress next ones by keeping note of their arrival faces, thus enabling a native multicast dissemination model.
Accordingly, three data structures are used in NDN: the Content Store (CS), representing the cache memory; the Forwaring Information Base (FIB) that plays the same role as in classic IP routers; and the Pending Interest table (PIT) in which it is possible to keep track of the arrival faces of the \textit{Interest} messages received in the past and for which a \textit{Data} packet has not been yet received.

\section{Overlay Service Capability Layer}\label{distributed}
This paper proposes an OSCL to increase scalability, robustness, and  flexibility of ETSI-M2M systems. From a networking perspective, the OSCL is a meshed overlay that inter-connects many SCL instances to allow the provisioning of M2M services without the compulsory inter-mediation of the NSCL. 

In the ESTI M2M standard, all devices and gateways should be mutually authenticated at the NSCL. In absence of mobility, this operation is executed only once per device and allows the M2M system to handle future resource discovery operations. This kind of discovery is centralized and it uses the NSCL as a relaying point of all signaling messages. On the opposite, in our proposal, it will be used as a fallback solution only when the OSCL fails to handle a discovery operation.

Similarly, after a resource has been discovered, the NSCL is in charge to relay subscriptions to that resource (e.g., the sensing service of an electricity metering application) and all the data sent to subscribers (e.g., a remote energy monitoring application). Contrariwise, we propose to leverage OSCL capabilities to access to resources in a distributed way.


In the ETSI M2M standard, the resources are named according to a hierarchical scheme. To provide an example, a possible resource name for the last reading of a sensor, in the ETSI M2M standard, could be:

\url{Gscl1/applications/meter_app/containers/meter_data/content_instances/latest}.

Accordingly, we propose to adopt the NDN architecture to handle the services offered by the OSCL. Thanks to this choice a named resource can be located by simply issuing an \textit{Interest} message containing its name. This message will be routed within the OSCL under the control of the NDN strategy layer and it will trigger at least one \textit{Data} message containing the locator of the target resource. Many routing algorithms are nowadays available for NDN based on different kind of assumptions \cite{Xylomenos:2013}.
The choice is application specific and can be left open in order to allow customized M2M deployments. It is worth to remark that  \textit{Interest} routing operations could fail for many reasons (e.g., stale information in the FIB of the nodes, resource not reachable due to a not connected topology graph, research scope too limited to reach the target resource). Under such circumstances, the classic centralized resource discovery already available in the ETSI M2M standard is used as a backup solution.

After the resource discovery has been accomplished, a \textit{QoS Monitoring} phase is executed. If the resource was discovered using the distributed approach, then a path in the OSCL already exists that could be used for subsequent subscriptions and data exchange phases. The \textit{QoS Monitoring} phase will measure the key performance indicators of the path (e.g., packet loss ratio, throughput, delay) and if the quality of the path will result to be poor a new link will be established in the OSCL to enable a direct access to the target resource. The link will be set up using the locator information provided by the discovery phase. Of course, if the distributed discovery failed and, as a consequence, the centralized one was used, a new link will be immediately set up in the OSCL. We remark that, similarly to \cite{Li:2011}, we extend the \textit{Interest} header, and related in network handling operations, in order to solicit the transmission of more than one \textit{Data} message by the target resource. 

\begin{figure}[htbp]
	\centering
		\includegraphics[width=7 cm]{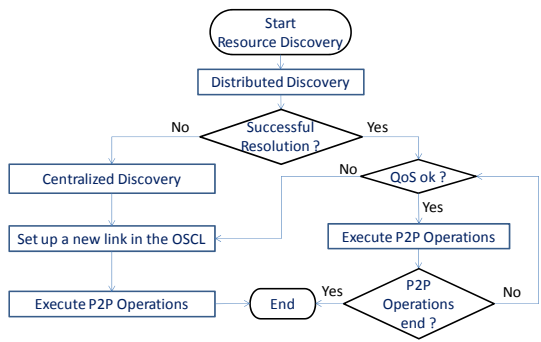}
	\caption{OSCL operations: simplified flow chart.}
	\label{flow-chart}	
\end{figure}

In summary, the Overlay Service Capability Layer (OSCL) enables the following enhancements (see also Fig. \ref{flow-chart}): 

\textbf{Distributed Service Discovery}. It is used to discover a resource in the M2M system based on ICN primitives. This phase, if successful, opens at least one path in the OSCL towards the discovered resource. Also, it returns both the URI of the resource and its network locator, e.g., IP address and Port. These two additional information are useful to handle alias and to establish a new link in the OSCL (if required), respectively. 

\textbf{P2P Subscription \& Data Exchange}. Thanks to the discovery phase, subscriptions to a target resource and subsequent exchange of information can be handled without the centralized control of the NSCL, i.e., a multihop path in the OSCL can be used. 

\textbf{QoS Monitoring}. The path activated using the \textit{Distributed Service Discovery} is composed by a sequence of logical links and nodes within the OSCL. It can be used to access to the remote resource. At the same time, the QoS this path is able to provide could not be enough for effectively supporting \textit{P2P Subscription \& Data Exchange} phases. As a matter of fact, some nodes within the overlay could not be able to effectively act as network relayers and thus worsening the QoS due to additional delays and packet losses. For this reason, the OSCL we envisage also include a \textit{Distributed QoS Monitoring} service that establish a new link with the remote resource when the path being used is not suitable anymore for a given M2M application.   

Notice that all these new features are optional so that the legacy interoperability with existing ETSI M2M deployments would be ensured together with the possibility of a graceful upgrade of currently available systems.

\section{Example Use Cases}\label{usecases}
In this section we describe two basic uses cases to demonstrate the main features of the proposed approach. The first one is made of the M2M Network, one gateway, and one device (see Fig. \ref{etsi-m2m-example1}). The gateway and the device host an electricity metering application and an energy monitoring application, respectively. The monitoring application needs to read the data sensed by the metering process. Before this becomes possible, \underline{if OSCL is not used}, it is necessary to accomplish the following actions: (1) the gateway and the device mutually authenticate to the NSCL; (2) the electricity metering application registers to the GSCL and then creates a data container.; (3) the monitoring application registers to the DSCL; (4) the monitoring application discovers the name of the gateway by querying the NSCL and then discovers the name of the application by querying the GSCL. Both queries are routed by the NSCL. At the end of this process the monitoring application knows the URI of the metering application (e.g., \url{GSCL/applications/electricity_meter}); (5) the monitoring application issues a subscription targeted to that URI. This subscription is routed by the NSCL towards the GSCL; (6) after the GSCL accepts the subscription, every time the electricity metering application pushes a new event to the data container (i.e., a new measurement) a notification is sent to the monitoring application (through the NSCL).

\begin{figure}[htbp]
	\centering
		\includegraphics[width=7 cm]{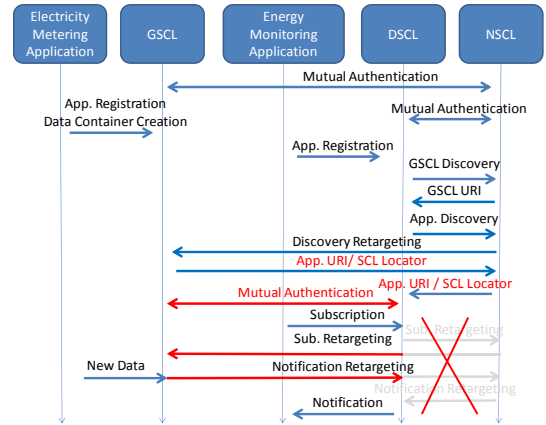}
	\caption{Message Sequence Chart with OSCL. Straight lines represent the ETSI M2M messages used also by the OSCL. Dashed lines represent the new messages introduced with the OSCL. The gray lines are the ETSI M2M messages that are no longer used thanks to the OSCL.}
	\label{sc2}	
\end{figure}

\underline{By applying the OSCL} to this use case, it becomes possible to increase the scalability of the ETSI M2M system. In fact, as shown in Fig. \ref{sc2}, the NSCL is no longer in charge of relaying the data sensed from the metering application. 

\begin{figure}[htbp]
	\centering
		\includegraphics[width=8 cm]{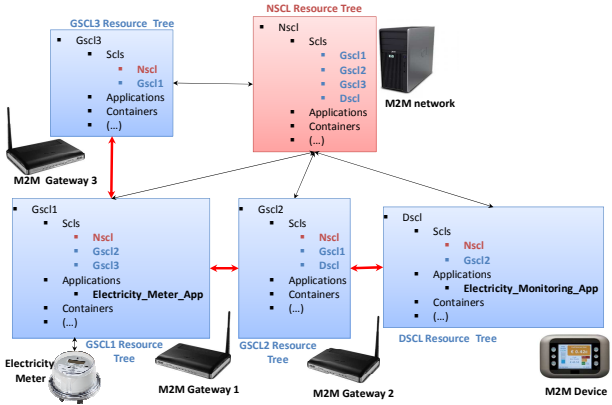}
	\caption{An extended ETSI M2M scenario with P2P capabilities.}
	\label{scenario2}	
\end{figure}

Also, from the message sequence chart in Fig. \ref{sc2}, it is possible to note as the authentication, discovery, subscription, and data exchange operations do not require significant modifications to ETSI messages: they only entail different endpoints in the exchange of packets. 
A slightly more complex scenario is depicted in Fig. \ref{scenario2}, where two more gateways are present with respect to the previous case. This second scenario allows us to detail how NDN messages (i.e., \textit{Interest} and \textit{Data}) are routed through multiple GSCL instances in the OSCL. 

\begin{figure}[htbp]
	\centering
		\includegraphics[width=7 cm]{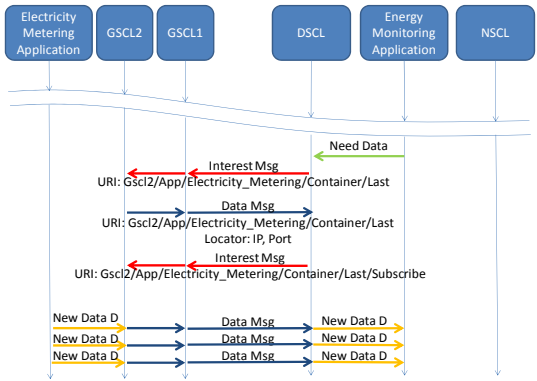}
	\caption{Message Sequence Chart with OSCL and ICN messages.}
	\label{sc3}	
\end{figure}

Fig. \ref{sc3} shows that the \textit{Interest} message opens a path towards the target resource during the discovery phase. This path is then used to transport both signaling and application data (encapsulated in \textit{Data} messages) between two remote SCL instances in the OSCL, without any mandatory intervention of the NSCL. It is also worth to note that relaying data through a multi-hop path reduces the number of logical links established in the overlay per each SCL instance thus further increasing the scalability of the system.

\section{Performance Evaluation}
To provide a preliminary assessment of the properties of the OSCL, we have considered a complex scenario composed by $N$ SCL instances (with $N$ ranging from $2^5$ to $2^{16}$) and assigned to the \textit{QoS Monitoring} function the role of not allowing the creation of paths longer than $D$ hops (with $D$ ranging from $3$ to $10$). This requirement is particularly relevant in order to shorten the maximum packet delay in the data sessions. To this end, as soon as a new couple of nodes needs to communicate, a straight link in the OSCL is created among them, \underline{if and only if} no path already exists shorter than $D$ hops. In particular, we have considered a long random sequence of couples of nodes wishing to establish a data session and we observed how the degree (i.e., average number of links per node) of the OSCL evolves over the time, subject to the constraint imposed by the \textit{QoS Monitoring} function. At the present stage, we have not considered the impact of caching because it will affect the average delay and load rather then the properties of the topology. Research in this direction is in progress. Based on this scenario, results indicate that the OSCL average degree can be approximated by $O(\sqrt[D]{2N \ln N})$. This expression has been derived using  graph theoretical arguments \cite{Loguinov05} and validated using computer simulations (to this end we devised a system level simulator in Matlab\footnote{\url{http://www.mathworks.com/}}). It provides a way to predict the relaying overhead of SCL instances, which is closely coupled to the number of links per node in the OSCL topology, and to finely tune $D$ based on the scale ($N$) of the M2M system and the expected capabilities of its components.


\section{Conclusion and Future Research}
An ICN approach to M2M systems has been presented in this paper in order to enhance the ETSI M2M standard. 
As future works we plan to: (i) evaluate the performance of the resulting OSCL in realistic settings; (ii) afford the problem of an efficient name space design; (iii) investigate topology optimization mechanisms; (iv) add the possibility to express goals in content names.

\section*{Acknowledgments}
This work was supported by the INSA of Toulouse (FR) and the PON project RES NOVAE, funded by the Italian MIUR and by the European Union (European Social Fund).

\begin{singlespace}
\bibliographystyle{IEEEtran}
\bibliography{biblio}
\end{singlespace}

\end{document}